\documentclass[twoside]{article}
\usepackage{graphicx}
\usepackage{fancyhdr}
\usepackage{multicol}
\usepackage{styl-ap}

\begin{document}
\title{Broad-band variability in accreting compact objects}
\author{Simone Scaringi\work{1,2}}
\workplace{Instituut voor Sterrenkunde, K.U. Leuven, Celestijnenlaan 200D, Leuven, Belgium
\next
Max Planck Institute fur Extraterrestriche Physik, D-85748 Garching, Germany}
\mainauthor{simo@mpe.mpg.de}
\maketitle

\begin{abstract}
Cataclysmic variable stars are in many ways similar to X-ray binaries. Both types of systems possess an accretion disk, which in most cases can reach the surface (or event horizon) of the central compact object. The main difference is that the embedded gravitational potential well in X-ray binaries is much deeper than those found in cataclysmic variables. As a result, X-ray binaries emit most of their radiation at X-ray wavelengths, as opposed to cataclysmic variables which emit mostly at optical/ultraviolet wavelengths. Both types of systems display aperiodic broad-band variability which can be associated to the accretion disk. Here, the properties of the observed X-ray variability in XRBs are compared to those observed at optical wavelengths in CVs. In most cases the variability properties of both types of systems are qualitatively similar once the relevant timescales associated with the inner accretion disk regions have been taken into account. The similarities include the observed power spectral density shapes, the rms-flux relation as well as Fourier-dependant time lags. Here a brief overview on these similarities is given, placing them in the context of the fluctuating accretion disk model which seeks to reproduce the observed variability. 
\end{abstract}


\begin{multicols}{2}
\section{Introduction}
Cataclysmic variables (CVs) are close interacting binary systems where a late-type star transfers material to a white dwarf (WD) companion via Roche lobe overflow. With an orbital period ranging from hours to minutes, the transferred material from the secondary star forms an accretion disc surrounding the WD. As angular momentum is transported outwards in the disc, material will approach the innermost regions close to the WD in the absence of strong magnetic fields, and eventually accrete on to the compact object. X-ray binaries (XRBs) are also compact interacting binaries which are similar to CVs in many ways, but where the accreting compact object is either a black hole (BH) or a neutron star (NS). Both CVs and XRBs, as well as active galactic nuclei (AGN; accreting supermassive BHs), have been shown to display strong aperiodic variability on a broad range of time-scales as well as in different wavelength ranges.

XRBs have shown variability ranging from milliseconds to hours, whilst for CVs this ranges from seconds to days. This difference can be mainly attributed to the fact that the innermost edges of the accretion discs in CVs sit at a few thousand gravitational radii, whilst for XRBs it can reach down to just a few gravitational radii. The fact that material can get deeper within the gravitational potential of XRBs, as compared to CVs, also explains why they are more luminous and emit predominantly in X-rays, compared to CVs, which emit predominantly at optical/UV wavelengths. Aperiodic broad-band variability (also referred to as flickering) has extensively been studied in X-rays for XRBs over several decades in temporal frequency (see for example Terrell 1972; van der Klis 1995; Belloni et al. 2000; Homan et al. 2001; Belloni, Psaltis \& van der Klis 2002). As CVs emit mostly at optical/UV wavelengths, timing studies of these objects had to rely on optical observing campaigns from Earth, which are inevitably hindered by large interruptions, poor cadence, and in many cases poor signal-to-noise ratios. Furthermore, the key time-scales to probe in CVs are much longer than in XRBs, requiring long, uninterrupted observations. Recently, CV timing studies have been facilitated thanks to the advent of the \textit{Kepler} satellite (Gilliland et al. 2010; Jenkins et al. 2010), which is able to provide long, uninterrupted and high-precision light curves in the optical light from space. Thanks to these capabilities it is now possible to probe over four orders of magnitude in temporal frequency in CVs. More importantly, it is now possible to compare the aperiodic variability properties observed in XRBs to those observed in CVs after taking into account the relevant timescale and wavelength ``translations''. 

\section{Broad-band aperiodic variability in accreting white dwarfs}
One important discovery relating the flickering properties observed in CVs to those observed in XRBs has come from studying the \textit{Kepler} lightcurve of the nova-like CV MV Lyrae displaying typical flickering behaviour as observed in other CVs. Specifically, Scaringi et al. 2012a have reported the discovery of the so-called rms(root-mean-square)-flux relation within the \textit{Kepler} lightcurve of MV Lyrae, where the rms-variability power linearly correlates with the mean cont rate of the source. This relation has been previously observed in a number of XRBs and AGN (Uttley \& McHardy 2001; Uttley et al. 2005), and, together with the observed log-normal flux distributions, rules out simple additive processes as the source of flicker noise (e.g. superposition of many independent ‘shots’), and instead strongly favours multiplicative processes (e.g. coupling of mass-transfer variations travelling from the outer to inner disc for the latter) as the source of variability. More importantly, the fact that very similar rms-flux relations are found within all different types of accreting compact objects (BHs, NSs, WDs) on all scales strongly suggests that the driving mechanism responsible for the observed aperiodic variability is the same in all systems, irrespective of mass, size or type.

Additionally to displaying the rms-flux relation, Scaringi et al. 2012b have reported that the power spectral density (PSD) of MV Lyrae is also qualitatively similar to those observed at X-ray wavelengths in XRBs. Specifically, all PSDs display single or multiple quasi-periodic osscillations (QPOs) as well as a high frequency break. Both the PSDs of XRBs and CVs can be qualitatively modelled with a combination of Lorentzian shaped functions, with the main difference arising from the characteristic frequencies involved. For example, XRBs display aperiodic variability on a wide range of timescales with high-frequency breaks at $\approx10^{0}-10^{1}$ Hz (Belloni et al. 2005), whilst CVs display very similar PSD, but scaled to lower frequencies such that the high frequency break occurs at $\approx10^{-3}$ Hz. The difference between the PSDs in XRBs and CVs can be mainly attributed to the fact that material within the accretion disk can fall deeper within the embedded gravitational potential well of XRBs as opposed to CVs. This will result in XRBs displaying variability on shorter timescales than in CVs, and furthermore, will result in XRBs emitting most of their radiation at X-ray wavelengths as opposed to CVs emitting mostly at optical/UV wavelengths.

\section{Fourier-dependent time-lags in CVs}
Similarities between the flickering properties of XRBs and CVs are not only limited to single-band observations. It has been known for over a decade that XRBs display high levels of coherence between two simultaneously observed X-ray lightcurves in different energy bands (Vaughan \& Nowak 1997; Nowak et al. 1999). Associated to this, Fourier-dependent time-lags are also observed at X-ray wavelengths, where hard X-ray photons are delayed with respect to the soft photons, with larger delays at the lowest temporal frequencies (known as hard lags). A natural explanation for hard lags in XRBs comes from the fluctuating accretion disk model (see Arevalo \& Uttley 2006 and references therin), where the coupling of fluctuations in the mass-transfer rate at different accretion disk radii is responsible for the observed variability. As matter propagates from the cooler outer regions of the disk towards the hotter inner ones, the observed variability in two different energy bands will be delayed with respect to each other as a result of the temperature gradient 
in the accretion disk and fluctuations moving inwards.

Using the ULTRACAM instrument (Dhillon et al. 2007) mounted on the 4.2 meter William Herschel Telescope, Scaringi et al. 2013 have obtained simultaneous optical lightcurves in the $u'$, $g'$ and $r'$ bands on the two nova-like systems MV Lyrae and LU Cam. The results from this observing campaign show that both systems display soft lags (where blue photons are observed before the red ones) at the lowest observed frequencies, with larger lags at low frequencies. Soft-lags have also been observed at X-ray wavelengths in XRBs, as well as AGN, and have been explained as photoionising reflection of hard-X-ray photons originating from the inner regions of the disk (possibly the corona) onto the outer regions. In this case, the lag delay would be the light-travel time from the emitting region to the reflecting region. The soft lags observed in Scaringi et al. 2013 are however too large (on the order of $\approx10$ seconds for LU Cam) to be explained by light-travel time delay. Nevertheless, it might still be that photons from the 
hot inner regions are being reprocessed by the cooler outer disk, possibly on the thermal timescale.

\section{A physical model for the flickering variability}
The observed flickering properties in XRBs have been modelled in the context of the fluctuating accretion disk model (see Arevalo \& Uttley 2006 and references therin). Ingram \& Done (2011,2012) have shown how this model can be successfully applied to reproduce the observed PSD shapes in XRBs. In the most recent attempt by Ingram \& van der Klis (2013), the authors have demonstrated how the model can be implemented analytically (rather than numerically), thus reducing the computation time of specific models from hours to seconds. In Scaringi 2013 a simplification of the Ingram \& van der Klis (2013) model has been applied to the \textit{Kepler} lightcurve of the CV MV Lyrae by removing general relativistic effects which are negligible in accreting WDs. The model prescription employed allows to fit the high-frequency PSD of MV Lyrae by associating the observed flickering to the viscous timescale at specific disk radii. Assuming the accretion disk extends all the way to the WD surface, the model allows to fit for the disk outer edge, as well as for $\alpha(h/r)^2$, where $\alpha$ is the viscosity parameter and $h/r$ the disk scale height (Shakura \& Sunyaev 1973). A qualitatively good fit to the data is achieved (with reduced $\chi^2\approx1.2$). The model parameters resulting from the fit are displayed in Table~1, and suggest that the observed high-frequency flickering is driven by a geometrically thick disk extending from $r\approx0.12R_{\odot}$ all the way to the WD surface. 

\begin{mytable}
\caption{Best-fit parameters with associated ($1\sigma$) errors for the fluctuating accretion disk model applied to the \textit{Kepler} lightcurve of MV Lyrae (Scaringi 2013).}
\label{author-tab1}
\bigskip
\centerline{\begin{tabular}{|l l|}
\hline
$M_{WD}(M_{\odot})$    &    $\equiv0.73$       \\ 
$r_{in}(R_{\odot})$    &    $\equiv0.0105 $      \\   
$r_{out}(R_{\odot})$   &    $0.117^{+0.029}_{-0.020}$ \\  
$\gamma$               &    $0.853^{+0.047}_{-0.041}$  \\ 
$\alpha(h/r)^2$        &    $0.705^{+0.289}_{-0.182}$  \\ 
$F_{var}$              &    $0.220^{+0.001}_{-0.001}$ \\ 
\hline
\end{tabular}}
\end{mytable}

Although the observed emission at optical wavelengths in CVs is though to originate from a cold, geometrically thin outer disk, the results from this analysis seem to suggest that the flickering is driven by a geometrically thick inner disk. It is thus possible that the geometrically thin outer disk is reprocessing photons from the geometrically thick inner one, allowing to explain the inferred results. Similar geometric configurations have also been inferred in XRBs. Both X-ray timing and spectral analysis suggest that geometrically thick disk exists close to the compact object (optically thin, referred to as the corona), possibly sandwiching the geometrically thin, optically thick disk. If a similar configuration is confirmed in CVs, then it is possible that the physics responsible for generating the inner geometrically thick disks in CVs might be the same as that in XRBs.  

\section{Conclusion}
Both CVs and XRBs, as well as AGN, are observed to display aperiodic broad-band variability. This variability can be associated to the accretion disks in these systems. As the accretion disks in XRBs can fall much deeper within the embedded gravitational potential well as opposed to CVs, most of their emission will occur at X-ray wavelengths as opposed to optical/ultraviolet. Furthermore, the timescales associated with the disk inner edge are a few orders of magnitude higher in temporal frequency than those of CVs. Here, a brief overview of the broad-band variability properties observed in CVs has been presented. In particular, the aperiodic properties discussed (PSD shapes, rms-flux relations and Fourier-dependent time-lags) are in many ways similar to those observed at X-ray wavelengths in XRBs once the relevant temporal scaling has been taken into account. The fluctuating accretion disk model which seeks to explain the observed variability in XRBs has also been briefly discussed in the context of CVs. 
This model associates the observed variability to the viscous timescale at specific disk radii. In this respect, applying this model to the \textit{Kepler} lightcurve of the nova-like CV MV Lyrae suggests the existence of a geometrically thick disk close to the WD responsible for driving the observed high-frequency flickering. 

\thanks
The author wishes to acknowledge funding from the FWO Pegasus Marie-Curie fellowship program, as well as useful and insightful discussions with Christian Knigge, Elmar Koerding, Phil Uttley and Tom Maccarone.

\bigskip
\bigskip
\noindent {\bf DISCUSSION}

\bigskip
\noindent {\bf LINDA SCHMIDTOBREICK:} For NLs we see a difference in the spectra between low and high inclination systems. Low inclination systems show Balmer absorption, possibly from an inner optically thick accretion disk which might coincide with your geometrically thick inner disk. Does one see a similar relation of the presence of high frequency flickering with inclination?

\bigskip
\noindent {\bf SIMONE SCARINGI:} The current model has only been applied on one system at the moment (MV Lyrae, low inclination). Future applications of the model on other systems will be able to determine whether the high frequency flickering displays different properties as a function of inclination.

\bigskip
\noindent {\bf SOLEN BALMAN:} How do you get an optically thick disk out of $\alpha(h/r)^{2}=0.7$? If $\alpha=0.1$ the disk is hot and it is no longer optically thick! 

\bigskip
\noindent {\bf SIMONE SCARINGI:} The inference of $\alpha(h/r)^{2}=0.7$ from the modelling of the PSD in MV Lyrae does not suggest that the disk is optically thick. It might well be that the inner region of the disk inferred here are optically thin as suggested.

\bigskip
\noindent {\bf RAYMUNDO BAPTISTA:} Is you $\alpha$ constant with radius?

\bigskip
\noindent {\bf SIMONE SCARINGI:} In the current model prescription $\alpha$ is kept as a constant with radius. However, it is more physical and realistic to allow $\alpha$ to have a radial dependence. This will be explored with the current model in the future.

\end{multicols}
\end{document}